\documentclass[12pt]{article}
\pdfoutput=1
\usepackage{graphicx,psfrag,epsf,color}
\usepackage{amsmath,amssymb,amsfonts}
\usepackage{array}
\usepackage{cite}
\usepackage{rotating}
\usepackage{slashed, cancel}
\usepackage{xcolor}
\newcounter{MBQ}

\newcounter{RSQ}

\newcounter{MGQ}

\def\slash#1{#1 \hskip-0.45em /}

\newcommand{\be}{\begin{equation}}
\newcommand{\ee}{\end{equation}}
\newcommand{\bea}{\begin{eqnarray}}
\newcommand{\eea}{\end{eqnarray}}
\newcommand{\bi}{\begin{itemize}}
\newcommand{\ei}{\end{itemize}}
\newcommand{\ben}{\begin{enumerate}}
\newcommand{\een}{\end{enumerate}}
\newcommand{\bt}{\begin{tabular}}
\newcommand{\et}{\end{tabular}}

\newcommand{\nn}{\nonumber}

\newcommand{\T}{{\bf T}}

\newcommand{\nnm}[1]{n_{#1-}}
\newcommand{\nnp}[1]{n_{#1+}}

\begin{document}
\allowdisplaybreaks

\begin{titlepage}

\begin{flushright}
{\small
TUM-HEP-1210/19\\
July 11, 2019
}
\end{flushright}

\vskip1cm
\begin{center}
{\Large \bf\boldmath Violation of the Kluberg-Stern-Zuber 
theorem\\[0.1cm] in SCET}
\end{center}

\vspace{0.5cm}
\begin{center}
{\sc Martin~Beneke,$^a$ Mathias~Garny,$^a$ Robert~Szafron,$^a$ Jian~Wang$^{a,b}$} \\[6mm]
{\it $^a$Physik Department T31,\\
James-Franck-Stra\ss e~1, 
Technische Universit\"at M\"unchen,\\
D--85748 Garching, Germany\\
$^b$School of Physics, Shandong University, Jinan, Shandong 250100, China
}
\end{center}

\vspace{0.6cm}
\begin{abstract}
\vskip0.2cm\noindent
A classic result, originally due to Kluberg-Stern and Zuber, states 
that operators that vanish by the classical equation of motion (eom) 
do not mix into ``physical'' operators. Here we show that and 
explain why this result does not hold in soft-collinear effective 
theory (SCET) for the renormalization of power-suppressed  
operators. We calculate the non-vanishing mixing of eom operators 
for the simplest case of $N$-jet operators with a single collinear 
field in every direction. The result implies that---for the 
computation of the anomalous dimension but not for on-shell matrix 
elements---there exists a preferred set of fields that must be used 
to reproduce the infrared singularities of QCD scattering amplitudes. 
We identify these fields and explain 
their relation to the gauge-invariant SCET Lagrangian. Further checks 
reveal another generic property of SCET beyond leading power, which 
will be relevant to resummation at the next-to-leading logarithmic 
level, the divergence of convolution integrals with the hard matching 
coefficients. We propose an operator solution that allows to 
consistently renormalize such divergences.
\end{abstract}
\end{titlepage}

\section{Introduction}

Operators that vanish by the classical equation of motion (eom) are 
usually considered redundant in the construction of an operator basis 
for an effective field theory (EFT) because the EFT is only designed 
to reproduce the on-shell scattering amplitudes of the low-energy 
degrees of freedom.  
Indeed, eom operators do not affect on-shell S-matrix elements (see 
e.g.~\cite{Deans:1978wn, Politzer:1980me}), including infrared (IR) 
divergent matrix elements, provided the IR divergences are 
logarithmic~\cite{Deans:1978wn}. This is closely related to the fact 
that quantum fields provide a highly redundant representation of the 
S-matrix, which remains invariant under a large class of field 
redefinitions in writing the action of the theory.

Moreover, a well-known result due to Kluberg-Stern and Zuber 
\cite{KlubergStern:1975hc} states that eom operators (also termed 
``class~II'') do not mix into physical (``class~I'') operators under 
renormalization (see also \cite{Joglekar:1975nu,Deans:1978wn} and 
\cite{Espriu:1983zz,Collins:1994ee,Manohar:2018aog,Criado:2018sdb} for 
other aspects of the statement), 
which implies a triangular structure of 
the operator anomalous dimension matrix (ADM). This does not mean that 
eom operators can always be ignored in ADM computations. For example, 
if an off-shellness is employed to regulate IR divergences, or if, 
in the case of gauge theories, an IR regulator is used that breaks 
gauge invariance, counterterms proportional to eom operators may be 
necessary to remove subdivergences (see, for instance, 
\cite{Gambino:2003zm}). However, in the final ADM, the eom operators 
still mix only among themselves and do not influence the evolution of 
the couplings of the physical operators. Given the relation to field 
redefinitions, the triangular structure of the ADM is required by the 
consistency of the EFT since otherwise the ADM of the physical 
operators would depend on the arbitrary field representation 
employed to compute the S-matrix.

The scattering amplitudes of massless particles in gauge theories 
exhibit IR singularities due to the emission and loops of soft and collinear 
particles. In dimensional regularization (space-time dimension $d=4-
2\epsilon$), contrary to ultraviolet (UV) divergences, a double 
$1/\epsilon^2$ pole appears at the one-loop order, 
and up to $1/\epsilon^{2n}$ at the 
$n$th loop order. Referring to QCD in the following, the EFT that 
describes the high-energy scattering of massless partons (quarks and 
gluons) is soft-collinear effective theory 
(SCET) \cite{Bauer:2000yr,Bauer:2001yt,Beneke:2002ph,Beneke:2002ni}. 
The IR divergences of the QCD S-matrix appear as UV divergences 
of certain collinear field operators in SCET~\cite{Becher:2009cu}. 
Hence, the study of IR divergences and resummation of large IR 
logarithms in QCD can be phrased as an operator-mixing and 
renormalization-group problem in SCET.

This is well understood at leading power (LP) in the EFT expansion and 
has found many applications in collider physics. In this context, 
the expansion parameter of the EFT is defined as follows. Let $Q$ 
be the scale of a hard process that involves a number of jets. 
The typical transverse momentum $p_\perp\sim Q \lambda$ of partons 
within a jet defines 
the power counting parameter $\lambda \ll 1$ of SCET. The 
collinear modes within a jet can interact with other jets through the 
exchange of soft modes with momentum $k\sim Q\lambda^2$, 
which preserve the virtuality of collinear modes within jets. 
While the SCET Lagrangian beyond the leading power has been known 
for a long time \cite{Beneke:2002ph,Beneke:2002ni} and the first systematic 
study of power-suppressed effects and their factorization 
has been undertaken in the early days 
of SCET for the production of a single jet of hadrons in semi-leptonic 
$B$-meson decay \cite{Beneke:2004in}, a more complete investigation of
SCET beyond LP has started only recently  \cite{Larkoski:2014bxa,Freedman:2014uta,Kolodrubetz:2016uim,Feige:2017zci,Moult:2017rpl,Beneke:2017mmf,Beneke:2017ztn,Beneke:2018rbh,Moult:2018jjd,Beneke:2018gvs,Ebert:2018gsn}. 
In particular, the ADM of $N$-jet operators at next-to-leading power (NLP) 
$\mathcal{O}(\lambda,\lambda^2)$ has been computed at the 
one-loop order~\cite{Beneke:2017ztn,Beneke:2018rbh}.\footnote{An early computation 
of the matching coefficient and renormalization of a particular 
$\mathcal{O}(\lambda)$ suppressed operator relevant to exclusive and 
semi-inclusive $B$-meson decays can be found 
in \cite{Beneke:2004rc,Hill:2004if,Beneke:2005gs}.}

In this work, we show that the ADM of $N$-jet operators at sub-leading 
power in SCET violates the Kluber-Stern-Zuber (KSZ) theorem. In other words, 
we demonstrate that eom operators mix into physical operators. The 
mixing is in fact required to reproduce the IR divergences of QCD 
amplitudes from SCET. We explain where the proof of KSZ fails, 
why---contrary to the statements in the first two paragraphs of this 
introduction---SCET is nevertheless a sensible EFT of QCD. We discuss several checks.
They reveal another generic property of SCET beyond LP, which, although 
it has appeared in some applications \cite{Beneke:2017vpq,Alte:2018nbn}, has 
not yet been fully recognized: the divergence of 
the convolution integrals of hard coefficient functions with 
the SCET matrix element. We provide a rearrangement of the operator basis of 
$\mathcal{O}(\lambda)$ operators that can be consistently renormalized 
in the presence of singular convolutions with hard-matching coefficients. In the subsequent 
presentation we explain these issues, which are of interest from a 
general quantum-field theoretic point of view, in the simplest case. 
A complete treatment of SCET operator mixing at NLP including 
$\mathcal{O}(\lambda^2)$ mixing, as is relevant to NLP resummation 
at next-to-leading logarithmic accuracy, based on the techniques 
developed here is quite involved. We 
leave this to a longer and more technical paper \cite{inprep}.

\section{Lorentz invariance implies soft mixing}
\label{sec:cusp}

That something peculiar is going on with operator mixing in SCET 
beyond LP can be inferred from the following simple consideration. 
A LP $N$-jet operator is a product of $N$ collinear fields 
from $\{W_{k}^\dag\xi_{k},\, \bar{\xi}_k W_k, 
W_{k}^\dag [iD_{\perp k}^\mu W_{k}]\}\equiv \{\chi_k,\bar\chi_k,{\cal A}_{\perp k}^\mu\}$, representing the 
collinear-gauge-invariant 
collinear quark, antiquark and gluon fields, respectively, 
where $k=1,\ldots N$ specifies the well-separated directions of the 
$N$ jets.\footnote{Notation here and below 
as in \cite{Beneke:2017ztn,Beneke:2018rbh}.} The ADM of these 
operators up to the two-loop order has the form  
\begin{equation}
\mathbf{\Gamma} = -\gamma_{\rm cusp}(\alpha_s) \sum_{k<j} 
\mathbf{T}_k\cdot\mathbf{T}_j \ln\left(\frac{-s_{kj}}{\mu^2}\right) 
+ \sum_{k}
\gamma_k(\alpha_s) 
\label{eq:LPanomalousdim}
\end{equation}
in colour-operator notation \cite{Catani:1998bh} and for all out-going 
momenta $p_k$ at the operator vertex, 
 $s_{kj}=2 p_k\cdot p_j+i0$, $k,j=1\ldots N$. 
Here $\gamma_{\rm cusp}$ is the universal cusp anomalous dimension,
while $\gamma_k$ are the contributions to the anomalous dimension associated to each collinear direction $k$.
 For an $N$-particle 
scattering amplitude with exactly one line in every direction, it is natural 
to align the light-like vectors $\nnm{k}$ with the momentum 
$p_k$ such that $p_k^\mu = (\nnp{k} p_k) \,\nnm{k}^\mu/2$. In this 
case the above anomalous dimension is exact to all orders in the power 
expansion, in the sense that it 
generates the IR divergences of the full QCD amplitude, since operators with 
transverse derivatives have vanishing matrix elements.

However, it is equally possible to slightly misalign the reference 
vectors $\nnm{k}$ defining the collinear directions $k$, provided that 
the components of $p_k$ scale as  $(n_{k+}p,p_{\perp k},n_{k-}p)
\sim(1,\lambda,\lambda^2)$. Lorentz invariance implies that the anomalous 
dimension remains as in (\ref{eq:LPanomalousdim}), but the power 
expansion of SCET requires that it should be expanded for this 
situation. Defining 
\begin{eqnarray}
s^{(0)}_{kj}&=&\frac12(n_{k-}n_{j-}) (n_{k+}p_k)(n_{j+}p_j)\,,\\
s_{kj}&=& s^{(0)}_{kj}+ 
(n_{k+}p_k)(n_{k-}p_{\perp j})+(n_{j+}p_{j})(n_{j-}p_{\perp k})
 +\mathcal{O}(\lambda^2)\,,
\end{eqnarray}
we find
\begin{eqnarray}
\mathbf{\Gamma} &=& -\gamma_{\rm cusp}(\alpha_s) \sum_{k<j} 
\mathbf{T}_k\cdot\mathbf{T}_j \ln\left(\frac{-s_{kj}^{(0)}}{\mu^2}\right) 
+ \sum_{k}\gamma_k(\alpha_s) 
\nn\\
&&-\,\gamma_{\rm cusp}(\alpha_s) \sum_{k\not=j}
\mathbf{T}_k\cdot\mathbf{T}_j 
\frac{2 n_{j-}p_{\perp k}}{(n_{k-}n_{j-})(n_{k+}p_k)}
 +\mathcal{O}(\lambda^2)\,.
\label{eq:LPanomalousdimexp}
\end{eqnarray}
The $\mathcal{O}(\lambda)$ term in the second line implies non-vanishing 
operator mixing of the LP $N$-jet operator into NLP $N$-jet 
operators with an additional derivative $i\partial^\mu_{\perp k}$ acting on
one of the collinear building blocks.
The corresponding one-loop anomalous 
dimension in notation as in (2.17) of \cite{Beneke:2018rbh} is given by
\begin{equation}
\gamma^{kj}_{P_iQ_i} = 
 \frac{\alpha_s}{\pi}\,\T_k\cdot\T_j \,
\frac{n_{j-}^\mu\delta_{k i} }{(n_{k-}n_{j-})(n_{i+}p_i)} + 
(k \leftrightarrow j) \,,
\label{eq:NLPanomalousdim}
\end{equation}
where\footnote{The total anomalous dimension is obtained after summing 
over indices $k$ and $j$ with $k < j$. This leads to the requirement 
that $\gamma^{kj}_{P_iQ_{i}}$ must be symmetrized in $k$ and $j$. 
For $i\neq i'$ we find $\gamma^{kj}_{P_iQ_{i'}}=0$. 
} $P_i=J^{A0}_{1} \ldots J^{T1}_{i} \ldots J^{A0}_{N}$ is an $N$-jet operator
with LP building blocks $J^{A0}_{j}\in \{\chi_j,\bar\chi_j,{\cal A}_{\perp j}^\mu\}$
for $j\not=i$, and a time-ordered product with a  sub-leading power 
soft-collinear Lagrangian interaction $\mathcal{L}_i^{(1)}(x)$ at order $\lambda$,
\begin{equation}
J^{T1}_{i}(t_i) \equiv i\int d^4x \,
T\big\{J_i^{A0}(t_i),\mathcal{L}_i^{(1)}(x)\big\} \,,
\end{equation}
for $j=i$. Furthermore, 
$Q_i=J^{A0}_{1} \ldots J^{A1\,\mu}_{i} \ldots J^{A0}_{N}$ is an 
$\mathcal{O}(\lambda)$ suppressed current operator with\footnote{In the 
following we focus on a collinear quark in the $i$ direction
unless otherwise specified. In the latter case we explicitly indicate 
the collinear building block in the subscript, 
e.g. $J^{A1\,\mu}_{{\cal A}_{\perp i}^\nu}(t_i)=i\partial_{\perp i}^\mu{\cal A}_{\perp i}^\nu(t_i n_{i+})$ for a gluon building block.} 
$J^{A1\,\mu}_i(t_i)\equiv i\partial_{\perp i}^\mu J^{A0}_i(t_i n_{i+})$.
In the notation of \cite{Beneke:2017ztn,Beneke:2018rbh}, this corresponds to the operator mixing 
\begin{equation}
J^{T1}_{i}(t_i)  \to J^{A1\,\mu}_i(t_i)\,,
\end{equation}
through a soft loop. However, it was shown by explicit computation  
\cite{Beneke:2017ztn,Beneke:2018rbh} that no such soft mixing 
exists, in contradiction with the above conclusion. 

The correctness of (\ref{eq:NLPanomalousdim}) follows from another 
argument. Lorentz invariance is broken in SCET by the reference 
vectors $\nnm{k}$, but the implications of Lorentz invariance 
are instead encoded in reparameterization invariance (RPI) 
under changes of the arbitrary reference vectors \cite{Manohar:2002fd}. 
Without loss of generality (due to the dipole structure of the 
anomalous dimension at one loop), we consider $N=2$ and focus on 
the power-suppressed operators in the $i$-direction and $i< j$. RPI invariance implies 
that $J^{(A0,A0)}$ and  $J^{(A1,A0)}$ must appear in the 
combination
\begin{equation}
\int dsdt \,C^{(A0,A0)}(t,s)\,\bar{\chi}_j(s n_{j+}) \Gamma
\left[1+\frac{2 t}{n_{i-}n_{j-}}\,
n_{j-}\cdot\partial_{\perp i}\right]
\chi_i(t n_{i+})\,,
\label{eq:RPIA0A1}
\end{equation}
where $\Gamma$ is some Dirac matrix. In momentum space, the factor 
of $t$ is transformed into the derivative 
$(-i)\partial/\partial n_{i+} p_i$ acting 
on the momentum-space coefficient $C^{(A0,A0)}(n_{i+}p_i,n_{j+}p_j)$,
and the hard matching coefficients of  $J^{(A0,A0)}$ and  $J^{(A1,A0)}$ 
are therefore related as 
\begin{equation}
C^{(A1,A0)\, \mu}(n_{i+}p_i,n_{j+}p_j) = 
\frac{-2n_{j-}^\mu}{n_{i-} n_{j-}}\frac{\partial}{\partial n_{i+} p_i}\,
C^{(A0,A0)}(n_{i+}p_i,n_{j+}p_j)\,.
\label{eq:C1C0relation}
\end{equation}
This equation implies the renormalization group equation
\begin{eqnarray}
\frac{d}{d\ln\mu} \,C^{(A1,A0)\,\mu}(n_{i+}p_i,n_{j+}p_j) 
&=& -\bigg[\gamma_{\rm cusp}(\alpha_s) \T_i\cdot\T_j  
\ln\frac{-s_{ij}^{(0)}}{\mu^2} + \gamma_i
\bigg] C^{(A1,A0)\,\mu}(n_{i+}p_i,n_{j+}p_j) 
\nonumber\\
&& \hspace*{-3cm}+\,
\gamma_{\rm cusp}(\alpha_s) \T_i\cdot\T_j 
\frac{2n_{j-}^\mu}{(n_{i-}n_{j-}) n_{i+} p_i}\,
C^{(A0,A0)} (n_{i+}p_i,n_{j+}p_j)
\end{eqnarray}
for the coefficient function of the power-suppressed operator, as 
can be derived by applying $\partial/\partial n_{i+} p_i$ to the 
LP evolution equation. The inhomogeneous term implies $\mathcal{O}(\lambda)$ 
mixing into the $J^{(A1,A0)}$ operator with anomalous 
dimension given by~(\ref{eq:NLPanomalousdim}). 

By performing on-shell one-loop matching of the QCD amplitude 
$\langle \bar q(p_i) q(p_j)| J(0) | 0\rangle$ of the quark-antiquark 
current  $J(0)=\bar{\psi}\,\Gamma\psi$ to SCET, one can check that the 
counterterm related to the mixing anomalous dimension is indeed 
required to reproduce in SCET the IR divergence  of the on-shell QCD amplitude. 
Besides, one finds that this IR  divergence  appears in the soft region of the 
off-shell regulated QCD amplitude. We therefore conclude that there 
must exist soft mixing in SCET that is not captured by the conventional 
methods of computing operator mixing.

\section{Soft mixing from eom operators}
\label{sec:offshell}

\begin{figure}
\begin{center}
  \includegraphics[width=0.22\textwidth]{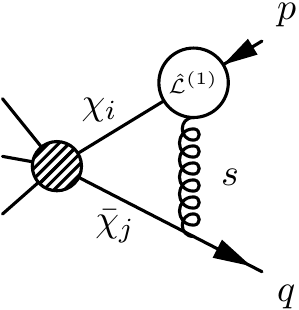}
\end{center}
\caption{\label{fig:soft_L1}
Soft loop with a single insertion of the sub-leading power Lagrangian $\hat {\cal L}^{(1)}_{\xi_i}$, that gives rise to mixing of the
time-ordered product $J^{T1}$ into $J^{A1}$. The three lines on the left symbolize possible further collinear directions, not present in our example.}
\end{figure}

Before we discuss eom operators, let us briefly review the computation of 
soft mixing in \cite{Beneke:2017ztn, Beneke:2018rbh}. Computations 
with the position-space SCET Lagrangian \cite{Beneke:2002ni,Beneke:2018rbh} 
yield a rather simple result for the soft one-loop integrals. At 
$\mathcal{O}(\lambda)$ there are only two relevant Lagrangian insertions,  
$ {\cal L}^{(1)}_{\xi_i}$ and 
$ {\cal L}^{(1)}_{\text{YM}_i}$.\footnote{The $\mathcal{O}(\lambda)$ 
suppressed coupling of soft quark fields to collinear fields is not 
relevant here, since it appears first at this order. The effect 
that we discuss in this section requires that there is a non-vanishing 
LP coupling.} 
In both cases, the structure of the soft-collinear coupling is the same, 
so in what follows we focus on
\be\label{eq:L1_gi}
 {\cal L}^{(1)}_{\xi_i} =
\bar{\xi_i} \left( x_{\perp i}^\mu n_{i -}^\nu \, W_i \,g_s F_{\mu\nu}^s 
W_i^\dagger \right) \frac{\slash n_{i+}}{2} \xi_i \,.
\ee 
The loop amplitude of the leading power operator with a single insertion 
of \eqref{eq:L1_gi} (see Figure~\ref{fig:soft_L1}) is  
\bea \label{eq:Amp_SCET}
2ig_s^{2} \, \T_i\cdot\T_j  \,\overline{u}_q \Gamma v_p \,
(n_{j+}q) (n_{i+}p ) \,\mu^{2\epsilon}\! 
\int \frac{d^dk}{(2\pi)^d}\frac{ (n_{j-} p_{\perp i}) 
(n_{i-}k)- (n_{i-}n_{j-})(p_{\perp i} k) }
{k^{2}\left(p^{2}-n_{i+}p \,n_{i-}k\right)^2\left(q^{2}+n_{j+}q \,n_{j-}k\right)}\,.\quad
\eea
Performing the tensor decomposition of the corresponding integral with 
numerator $k^\mu$, we note that the loop momentum $k$ in the numerator 
can be replaced by either $n_{i-}^\mu$ or $n_{j-}^\mu$. In the former case, 
the result is zero, since by definition $n^2_{i-}=0$ and 
$n_{i-}p_{\perp i} =0 $. In the latter case, the two terms in the numerator 
cancel. This implies that at  $\mathcal{O}(\lambda)$ there is 
no mixing of time-ordered product with the above Lagrangian into 
the $J^{A1}$ operator. 

The gauge-invariant form \eqref{eq:L1_gi} was obtained from the original 
SCET Lagrangian after a particular field redefinition. The original 
Lagrangian, which follows from the expansion of the QCD fields in the 
collinear and soft momentum region, reads \cite{Beneke:2002ph}
\be\label{eq:L1}
 \hat {\cal L}^{(1)}_{\xi_i} =
  \bar{\xi_i}\left(  i\slashed{D}_{\perp i}\frac{1}{in_{i+}D_i}g_s\slashed{A}_{s\perp i} 
  + g_s\slashed{A}_{s\perp i}\frac{1}{in_{i+}D_i} i\slashed{D}_{\perp i}  
  + [(x_{\perp i}\partial)(g_sn_{i-}A_s)]\right) \frac{\slash n_{i+}}{2} \xi_i\,.  
\ee
Here, (above) and below all soft fields $A_s$ without explicit
argument are evaluated at $x_{i-}^\mu=\frac12 n_{i+}x\, n_{i-}^\mu$, 
and $iD_i=i\partial+g_sA_{i}(x)+\frac12 g_s n_{i-}A_s \,n_{i+}$ where 
$A_i$ is the collinear gauge field for direction $i$. In the 
following, we focus on the abelian case to avoid unnecessary 
complications, and then return to 
non-abelian gauge symmetry in Section~\ref{sec:non-abelian}. For the 
abelian case the field redefinition reads explicitly 
\cite{Beneke:2002ni}
\be
\xi_i \to (1+ig_s x_{\perp i}A_s)\xi_i\,.
\ee
The difference between the Lagrangians \eqref{eq:L1} and 
\eqref{eq:L1_gi} is given by the eom Lagrangian
\bea
  \Delta{\cal L}^{(1)}_{\xi_i} \equiv \hat {\cal L}^{(1)}_{\xi_i}-{\cal L}^{(1)}_{\xi_i} &=&
  \bar{\xi_i}\left[ig_s x_{\perp i}A_s, in_{i-}D_i + i\slashed{D}_{\perp i}\frac{1}{in_{i+}D_i}i\slashed{D}_{\perp i} 
  \right] \frac{\slash n_{i+}}{2} \xi_i \nn\\
&=& \bar{\xi_i}\,ig_s x_{\perp i}A_s\frac{\delta S_{\xi_i}^{(0)}}
{\delta\bar\xi_i}+{\rm h.c.}\,,  
\label{eq:delL1}
\eea
where
\be
S_{\xi_i}^{(0)}=\int d^dx\,{\cal L}_{\xi_i}^{(0)}, \quad 
  {\cal L}_{\xi_i}^{(0)}=\bar{\xi_i}\left(in_{i-}D_i + i\slashed{D}_{\perp i}\frac{1}{in_{i+}D_i}i\slashed{D}_{\perp i}\right)\frac{\slash n_{i+}}{2} \xi_i
\ee
is the leading-power action. The eom Lagrangian 
$\Delta{\cal L}^{(1)}_{\xi_i}$  
does not contribute to on-shell matrix elements. For example, the diagram 
shown in Figure~\ref{fig:soft_L1} is proportional to $(p^2)^{-\epsilon}$ in 
$d$ dimensions. For the on-shell matrix element one performs the
limit $p^2\to 0$ \emph{before} taking the limit $\epsilon\to 0$. The rules 
of dimensional regularization then imply that $(p^2)^{-\epsilon}\to 0$,
in which case the eom Lagrangian gives rise to a scaleless integral 
that vanishes.

However, in order to extract the anomalous dimension from the 
UV divergences of the amplitude, one uses a small off-shellness of the external 
states. In practice, this means that one takes the limit $\epsilon\to 0$ 
first to extract the UV divergence, and then $p^2\to 0$, while the order of
limits is opposite when computing genuine on-shell matrix elements. 
The limit $p^2\to 0$ must exist for the anomalous dimension to be 
well-defined as has indeed been 
found~\cite{Becher:2009cu,Beneke:2017ztn,Beneke:2018rbh}.

We now consider the same diagram as before, but with insertion of 
$\hat {\cal L}^{(1)}_{\xi_i}$ from (\ref{eq:L1}), and find 
\bea\label{eq:integral_soft}
&& 2ig^{2} \,\T_i\cdot\T_j \,\overline{u}_q \Gamma v_p \,(n_{j+}q)(n_{i+}p )
\,\mu^{2\epsilon}\!
\int \frac{d^dk}{(2\pi)^d}\frac{1}{k^{2}\left(p^{2}-n_{i+}p\,n_{i-}k\right)^2\left(q^{2}+n_{j+}q \,n_{j-}k\right)}
\nonumber \\ 
&&\hspace*{0.5cm}\times 
\left[-(n_{j-} p_{\perp i})\frac{p^{2}}{n_{i+}p }
+\Big\{ (n_{j-} p_{\perp i}) (n_{i-}k)-
(n_{i-}n_{j-})(p_{\perp i} k)\Big\} \right]\nn \\  
&& =
\frac{\alpha_s}{\pi} \,\T_i\cdot\T_j \,\overline{u}_q \Gamma v_p\,
\frac{n_{j-}\cdot p_{\perp i}}{(n_{i-}n_{j-}) n_{i+}p}
\left(\frac{-\mu^{2}s^{(0)}_{ij}}{p^{2}q^{2}}\right)^{\!\epsilon}
\left(\frac{1}{\epsilon}+\mathcal{O}(\epsilon^0)\right).
\eea
The last two terms in the curly bracket in the second line reproduce the 
amplitude (\ref{eq:Amp_SCET}) obtained with the SCET Lagrangian insertion 
(\ref{eq:L1_gi}). The first term is, however, different. Naively, it would 
not be expected to contribute in the on-shell limit $p^2 \to 0$.  
However, the integral in the first line is proportional to 
$1/\epsilon\times 1/(p^2)^{1+\epsilon}$ (see \cite{Beneke:2018rbh}, Appendix~B). 
Therefore this term survives 
in the limit $p^2\to 0$ and contributes to the anomalous dimension if 
the expansion in $\epsilon$ is done first as is required in this case.  

The result~(\ref{eq:integral_soft}) contains a UV divergent 
contribution proportional to the $J^{A1}$  operator. To cancel 
it, we need the counterterm 
\be\label{eq:ZT1A1}
\delta Z^{kj}_{P_iQ_i} = \frac{\alpha_s}{2\pi \epsilon}
\, \T_k\cdot \T_j \,\frac{n_{j-}^\mu\delta_{ki}}{(n_{k-}n_{j-})n_{i+}p_i} 
+ (k\leftrightarrow j)\,.
\ee
In our specific example with $N=2$ we have 
$P_i=\bar{\chi}_{i'}\Gamma J^{T1}_{i}$ and $Q_i=\bar{\chi}_{i'}\Gamma J_i^{A1\,\mu}$
(where $i'=2$ for $i=1$ and vice versa) 
and after summing over all contributions $kj=12,21$ according to (2.14) 
in \cite{Beneke:2018rbh}, which gives a factor of two, 
we find that the counterterm (\ref{eq:ZT1A1}) 
cancels the divergence in (\ref{eq:integral_soft}).
The result holds also for the case of
open spin and color indices, which is straightforward in color operator 
notation, and does not get modified in the non-abelian case 
(see Section~\ref{sec:non-abelian}). 
Furthermore, it can be generalized to more than two jets ($N>2$), the only change being that $P_i$ and $Q_i$ are given by
the $N$-jet operators defined below \eqref{eq:NLPanomalousdim}.
Finally, the counterterm \eqref{eq:ZT1A1} yields an additional contribution 
when matching to the corresponding QCD current, that is precisely
of the form required to reproduce the IR divergences up to~$\mathcal{O}(\lambda)$. 

The  counterterm (\ref{eq:ZT1A1}) gives precisely the missing contribution 
(\ref{eq:NLPanomalousdim}) to the soft anomalous dimension.
Hence we have shown 
that, contrary to conventional wisdom, the eom operator (\ref{eq:delL1}), 
which produces the first term in~(\ref{eq:integral_soft}), mixes into 
physical operators. Moreover, it is the Lagrangian (\ref{eq:L1}) 
rather than (\ref{eq:L1_gi}) that reproduces the IR divergences  of the 
full QCD amplitude. This raises two fundamental questions: How is 
the KSZ theorem violated? What determines which Lagrangian or 
field representation of SCET is the one that correctly reproduces 
QCD in the IR, when employing an off-shell regulator to extract
the anomalous dimension?

\section{Violation of the Kluberg-Stern-Zuber 
theorem}

The fact that the renormalization of an eom operator (the part of the 
time-ordered product $J_i^{T1}$ containing $\Delta{\cal L}^{(1)}_{\xi_i}$) 
requires to introduce a counterterm proportional to a physical operator 
(here $J^{A1\,\mu}_i$) contradicts the statement that eom operators do 
not mix into physical operators. In the following, we summarize and adapt to 
our case the main points of the proof of this 
statement \cite{KlubergStern:1975hc}, which we refer to as the KSZ theorem.
We then show how one of the assumptions is violated in SCET.

Consider an action $S[\chi_i]$ that depends on a set of fields $\{\chi_i\}$.
A general eom operator can be written as
\be
\partial_S F(x) \equiv \int d^dy \,
\frac{\delta S}{\delta\chi_i(y)}K_i(y,x)F(x)\,,
\label{eq:eomOp}
\ee
where $F[\chi_i](x)\equiv F(x)$ is a local operator (e.g $\chi_1^2$ or 
$\chi_1\partial^\mu\chi_2$), and $K_i(y,x)$ are arbitrary c-number kernels 
(e.g. $\delta^{(d)}(x-y)$ or $\partial_\mu\delta^{(d)}(x-y)$). A summation over all 
types of fields $i$ is implied. In the following we leave this summation 
implicit, e.g. $J\chi\equiv J_i\chi_i$. The generating functional is
\be
Z[J,X,Y] = e^{i\,W[J,X,Y]} = \int {\cal D}\chi \,
e^{iS+i\int_x(J\chi+X\partial_S F+YF)}\,,
\ee
where sources $X(x)$ and $Y(x)$ are introduced to generate single insertions 
of $\partial_SF$ and $F$, respectively, and ${\cal D}\chi$ encompasses 
the integration over all fields. We also use the shorthand 
$ \int_x\equiv \int d^dx $.

\begin{figure}
\begin{center}
  \includegraphics[width=0.5\textwidth]{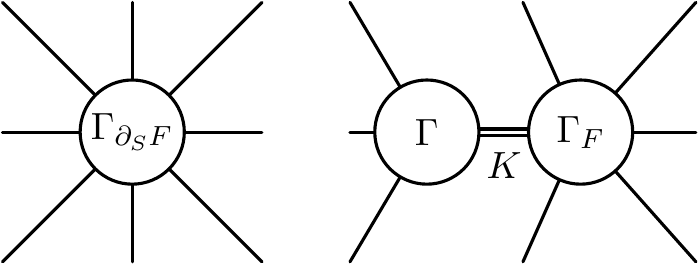}
\end{center}
\caption{\label{fig:kluberg}
Diagrammatic representation of \eqref{eq:1PI}. See text for details.}
\end{figure}

By performing the field redefinition $\chi(y)\to \chi'(y) = \chi(y) - \int_x K(y,x)F(x)X(x)$ and using ${\cal D}\chi={\cal D}\chi'$
one finds $Z[J,X,0] = Z[J,0,\bar Y[J,X]]+{\cal O}(X^2) $
where $\bar Y[J,X](x)=-\int_y J(y)K(y,x)X(x)$.
After taking a functional derivative and setting $X=0$, this yields the Ward identity for eom operators,
\be\label{eq:WardSource}
  \frac{\delta W}{\delta X(x)}\Big|_{X=Y=0} = -\int_yJ(y)K(y,x)\frac{\delta W}{\delta Y(x)}\Big|_{X=Y=0}\,.
\ee
For the one particle irreducible (1PI) effective action~\eqref{eq:WardSource} gives
\bea\label{eq:delGamma}
  \Gamma_{\partial_SF} &=& \int_y \left(\frac{\delta \Gamma}{\delta \chi(y)} K(y,x)\frac{\delta W}{\delta Y(x)}\right)\Big|_{X=Y=0}\nn\\
  &=& \int_y \left(\frac{\delta \Gamma}{\delta \chi(y)}\Big|_{X=Y=0}\right) K(y,x)\Gamma_{F}(x)
  \equiv \partial_\Gamma \Gamma_{F} \,.
\eea
This equation is still valid for arbitrary $\chi$, so we can generate 
relations for $n$-point, amputated, 1PI Green functions by taking derivatives 
with respect to $\chi$. Integrating over $x$ on both sides, assuming 
$K=K(y-x)$ (we depart from this choice later), and going to momentum space 
implies 
\bea\label{eq:1PI}
  \int_x \Gamma_{\partial_SF(x)}^{\rm 1PI}(p_1,\dots,p_n)  
  &=& \sum_{r=0}^n\sum_{\{i_1,\dots,i_r\}\atop\subset\{1,\dots,n\}}\int_p \Gamma^{\rm 1PI}(p_{i_1},\dots,p_{i_r},p)\nn\\
  && \times K(p)\times \Gamma_{F(p)}^{\rm 1PI}(p_{i_r+1},\dots,p_{i_n}),
\eea
where $\Gamma^{\rm 1PI}(p_1,\dots,p_n)$ are 1PI $n$-point functions,
and $\Gamma_{\cal O}^{\rm 1PI}(p_1,\dots,p_n)$ are 1PI $n$-point functions 
with an insertion of ${\cal O}$. Diagrammatically this implies factorization 
of the $n$-point function with insertion of the eom operator 
(\ref{eq:eomOp}) into a 1PI graph containing an insertion of $F$ and a usual 
1PI graph without any insertion, as illustrated in Figure~\ref{fig:kluberg}.
The left diagram shows the left-hand side of (\ref{eq:1PI}) for $n=8$, and 
the right one shows one contribution to the right-hand side for $r=3$. 
The blobs may contain an arbitrary number of loops.
As is apparent from Figure \ref{fig:kluberg}, the integration over $p$ does not count as
a loop: for $\Gamma^{\rm 1PI}(p_{i_1},\dots,p_{i_r},p)=
(2\pi)^d\delta^{(d)}(p_{i_1}+\cdots+p_{i_r}+p)
\tilde{\Gamma}^{\rm 1PI}(p_{i_1},\dots,p_{i_r},p)$
the integration is removed and $p=-(p_{i_1}+\cdots+p_{i_r})$ is fixed by 
momentum conservation. 

To prove the KSZ theorem, we proceed recursively in the number of loops.
We expand~\eqref{eq:delGamma} in the number of loops,
\bea
\Gamma_{\partial_SF}^{(L)} = \sum_{\ell=0}^L\partial_{\Gamma^{(\ell)}} 
\Gamma_{F}^{(L-\ell)}
\eea
and we assume that counterterms have been added to the action to make all 
1PI functions finite at loop order $0,\dots,L-1$. Then the remaining divergences 
at the $L$-loop order are
\bea
\Gamma_{\partial_SF}^{(L),\rm div} = \partial_{\Gamma^{(0)}} 
\Gamma_{F}^{(L),\rm div}+\partial_{\Gamma^{(L),\rm div}} \Gamma_{F}^{(0)}\, . 
\eea
Further, we assume that $K\times F$ contains a suppression in the power-counting 
parameter $\lambda$ (or in general, by some EFT power counting parameter 
$1/\Lambda$). Then, $\Gamma$ has to contain at least one power less than 
$\Gamma_{\partial_SF}$. Therefore, we can assume (in an extra recursion in 
powers of $\lambda$, or $1/\Lambda$) that $\Gamma^{(L)}$ is already 
rendered finite. Hence, only the first term contributes, and $\Gamma^{(0)}=S$ 
implies that
\bea
\Gamma_{\partial_SF}^{(L),\rm div} = 
\partial_{S} \Gamma_{F}^{(L), \rm div} \, .
\eea
Consequently, we can remove the $L$-loop divergences by adding to the action 
a term of the form $\sum_{F'}Z_{FF'}^{(L)}\partial_S F'$.
Here $S$ already contains counterterms up to order $L-1$.
If one \emph{assumes} that $\Gamma_{F}^{(L),div}$ and $K(p)$ are 
polynomials in the momenta (i.e. local divergences), then $\partial_{S} 
\Gamma_{F}^{(L),div}$ is an eom operator.
It follows that only eom operators $\partial_S F'$ are required to 
renormalize 1PI Green functions with a single insertion of $\partial_SF$, 
which proves the theorem.


Now we return to the question of why the KSZ theorem is violated for the
soft mixing computation in SCET. In the case of interest the relevant 
eom operator corresponds to $K^\mu(y,x)=-ix_{\perp i}^\mu\delta^{(d)}(y-x)$ and
\be\label{eq:Fexample}
F^\mu(x)=T\left\{J_i^{A0},g_s A_s^\mu(x_{i-})\xi_i(x)\right\}
\ee
(together with the corresponding hermitian-conjugated terms), which means 
that, due to the factor $x_{\perp i}^\mu$ in the kernel, a 
\emph{derivative} of the momentum-space 1PI Green function appears 
in~(\ref{eq:1PI}), which now reads
\bea
\int_x \Gamma_{\partial_SF(x)}^{\rm 1PI}(p_1,\dots,p_n)  
&=& \sum_{r=0}^n\sum_{\{i_1,\dots,i_r\}\atop\subset\{1,\dots,n\}}
\int_p \Gamma^{\rm 1PI}(p_{i_1},\dots,p_{i_r},p)
\nn\\
&& \times \frac{\partial}{\partial p_{\perp i}^\mu} 
\Gamma_{F^\mu(p)}^{\rm 1PI}(p_{i_r+1},\dots,p_{i_n}) \,.
\label{eq:idSCET}
\eea
In the example with two collinear directions $i$, $j$ 
discussed in the previous section, $n=2$, and the only contribution comes 
from $r=1$, because there is no one-point function generated by 
$\Gamma^{\rm 1PI}$. The above identity then takes the form
\bea
  \int_x \Gamma_{\partial_SF(x)}^{\rm 1PI}(p,q)  
  &=& \int_{\tilde{p}} \Gamma^{\rm 1PI}(p,\tilde{p}) \frac{\partial}{\partial \tilde{p}_{\perp i}^\mu} \Gamma_{F^\mu(\tilde{p})}^{\rm 1PI}(q).
\label{eq:idSCET2}
  \eea
To obtain the contribution relevant to the example at order $\alpha_s$, we 
need to insert the two-point tree-level function 
$\Gamma^{\rm 1PI}_{\rm tree}(p,\tilde p)$ and the one-point one-loop 
function with an insertion of (\ref{eq:Fexample}),
$\Gamma_{F^\mu(p)}^{\rm 1PI,1-loop}(q)$, whose explicit forms 
are\footnote{Omitting the Dirac structure, which is trivial, since 
the relevant vertices are spin-independent.}
\bea
\Gamma^{\rm 1PI}_{\rm tree}(p,\tilde{p} )&=&\frac{p^2}{n_{i+}p}
\,(2\pi)^d\delta^{(d)}(p-\tilde{p})\,,
\nn\\
\Gamma_{F^\mu(\tilde{p} )}^{\rm 1PI,1- \rm loop}(q)&=&
-\frac{\alpha_s}{4\pi}\frac{1}{\epsilon^2}\,
{\bf T}_i\cdot{\bf T}_j
\left(\frac{n_{i+} \tilde{p} 
n_{j+}q\,(n_{i-}n_{j-})\,\mu^2}{-2 \tilde{p}^2q^2}\right)^{\!\epsilon}
\frac{2n_{j-}^\mu}{(n_{i-}n_{j-})}\,.
\eea
The integral over $\tilde p$ is trivial in this case and after 
expanding 
in $\epsilon$ we find a divergent part that can only be renormalized by 
introducing a counterterm proportional to the physical operator 
$J_i^{A1\,\mu}$, in line with the previous explicit computation. 
Indeed, the divergent $1/\epsilon$ part of~(\ref{eq:idSCET2}) takes 
the form
\be
  \int_x \Gamma_{\partial_SF(x)}^{\rm 1PI, div}(p,q) \propto  \frac{p^2}{n_{i+}p} \times \frac{\partial}{\partial  p_{\perp i}^\mu} \underbrace{\Gamma_{F^\mu( p)}^{\rm 1PI, div}(q)}_{\propto(p^2)^{-\epsilon}/\epsilon^2} 
\ee
at the one-loop order. The first factor $p^2$ on the right-hand side 
corresponds to the contribution from $\partial_S$ in 
$\Gamma_{\partial_S F}^{\rm div}= \partial_S\Gamma_F^{\rm div}$, 
that---in the familiar cases, such as QCD--- would ensure 
that only eom counterterms are required to renormalize the eom operator. 
However, since in SCET $\Gamma_F^{\rm div}\propto \log(p^2)/\epsilon$ and 
$K(p)\sim \partial/\partial p_\perp$ are \emph{not} polynomials in the 
momenta, the assumption of locality in the KSZ proof is violated, 
and counterterms proportional to physical operators may be required. 
We conclude that the KSZ theorem is violated due to the momentum derivative coming from 
the $x_\perp^\mu$ term in the Lagrangian, which in turn arises from the 
multipole expansion, and because $\Gamma_F^{\rm div}$ has a
\emph{logarithmic} dependence on $p^2$ instead of polynomial dependence 
since in SCET double poles in $1/\epsilon$ appear already at one loop. 

\section{Uniqueness and preferred fields}

The violation of the KSZ theorem raises the question, which eom operators 
should be included in the derivation of the SCET anomalous dimensions. For 
example, in the usual situation any multiple of 
$\Delta{\cal L}^{(1)}_{\xi_i}$ could be added to the Lagrangian. In 
SCET, however, while not affecting the bare on-shell matrix elements, this 
would change the anomalous dimension by an arbitrary amount. 
In other words, there exists a preferred 
set of SCET fields which gives the correct anomalous dimension and 
reproduces the IR singular behaviour of QCD. Apparently, the 
preferred set of fields is the one in which the Lagrangian 
takes the form (\ref{eq:L1}), despite not being manifestly 
gauge-invariant.

The Lagrangian (\ref{eq:L1}) 
is obtained from full QCD by separating fields into 
collinear and soft, and by performing the multipole expansion of the 
soft fields without using the equations of motion \cite{Beneke:2002ph}. 
It therefore faithfully and directly reproduces the diagrammatic expansion in 
the method-of-region sense \cite{Beneke:1997zp} of the QCD amplitudes 
in the collinear and soft regions on- or off-shell. The preferred 
set of fields is therefore these ``original fields'' that construct 
the soft-collinear expansion of the {\em off-shell} QCD amplitudes. 
Since the SCET Lagrangian is not renormalized in any order of 
perturbation theory \cite{Beneke:2002ph}, the corresponding 
Lagrangian can easily be constructed to any desired order in 
the expansion in $\lambda$.

To sum up this discussion, the violation of the KSZ theorem implies 
that indeed there is a preferred set of fields to compute the 
anomalous dimensions of power-suppressed $N$-jet operators in SCET, 
at least when the UV singularities are extracted by calculating 
the amplitudes with off-shell IR regulators, as is usually done. 
The fields are uniquely fixed by the requirement to reproduce the 
off-shell amplitudes of QCD at tree-level due to the 
non-renormalization of SCET. 
The mixing of the eom operators into physical 
operators generates additional counterterm vertices proportional 
to the physical operators that must be taken into account in 
SCET calculations (see Section~\ref{sec:collemission} and 
Figure~\ref{fig:soft_qqg_FG1} below). However, once the counterterms 
and anomalous dimensions are determined, eom operators are no longer 
needed, field redefinition may be performed, and any Lagrangian 
(for instance, either (\ref{eq:L1_gi}) 
or (\ref{eq:L1})) obtained in this way can 
be used to compute {\em on-shell} amplitudes in SCET, since the 
insertion of eom operators into on-shell amplitudes gives 
scaleless integrals. Only the tree-level diagrams with the above-determined 
counterterm vertices are not scaleless, and must be included, 
in order to reproduce correctly the IR singularities of the 
full QCD amplitude. 

\section{Extension of soft mixing to the 
non-abelian case}
\label{sec:non-abelian}

It is quite cumbersome to write down the Lagrangian and eom terms in 
the ``original fields'' for the non-abelian theory including the Yang-Mills 
Lagrangian, and, in fact, this has never been done. In this section, we devise 
a method that side-steps this issue. It is enough to know the field 
redefinition that connects the fields in the gauge-invariant form 
of the Lagrangian to the ``original fields''. Both the gauge-invariant 
Lagrangian and field redefinition can be constructed from 
\cite{Beneke:2002ni}. This method allows us to efficiently compute 
the additional eom contributions to soft mixing, also at 
$\mathcal{O}(\lambda^2)$ \cite{inprep}.

We discriminate between two sets of field variables for each collinear 
direction: the original collinear quark $\hat\xi$ and gluon $\hat A_c$ fields 
that transform under soft gauge transformations as
\be
   \hat \xi \to U_s(x) \hat\xi, \quad \hat A_c\to U_s(x)\hat A_cU_s(x)^\dag
\ee 
and the redefined fields\footnote{Note that compared to 
\cite{Beneke:2002ni}, the hatted and un-hatted fields are interchanged.} 
$\xi$ and $ A_c$, which transform homogeneously in $\lambda$ as
\be
 \xi \to U_s(x_-)\xi, \quad 
 A_c\to U_s(x_-) A_cU_s(x_-)^\dag \,.
\ee 
For quantities expressed in terms of the fields $\xi$ and $ A_c$  
soft gauge invariance is manifest at every order in $\lambda$.
When computing \emph{on-shell} S-matrix elements, both sets of fields 
can be used. Indeed, the field operators are auxiliary quantities, and 
physical S-matrix elements do not depend on field redefinitions. 
Nevertheless, as seen above, this is not the case when keeping a (small) 
off-shellness on the external lines in order to extract UV divergences. 
We refer to these quantities as \emph{off-shell regulated S-matrix 
elements}.

From the practical point of view, it is desirable to use the redefined 
fields, such that the sub-leading power Lagrangian as well as the $N$-jet 
operator basis, are manifestly gauge invariant at every order in 
$\lambda$ \cite{Beneke:2002ni,Beneke:2017mmf}. On the other hand, to 
extract the anomalous dimension, the off-shell regulator requires the 
use of Green functions generated from the original fields. We now derive 
a relation that allows us to calculate the latter in terms of special 
Green functions of redefined fields.

We can express the original fields in terms of the 
redefined fields through relations of the form 
\bea
\hat \xi&=&F[\xi, A_c]=\xi+\lambda F^{(1)}[\xi, A_c]
+\lambda^2 F^{(2)}[\xi, A_c]+{\cal O}(\lambda^3)\,,
\nn\\
 \hat  A_c&=&G[\xi, A_c]= A_c+\lambda G^{(1)}[\xi, A_c]
+\lambda^2 G^{(2)}[\xi, A_c]+{\cal O}(\lambda^3)\,,
\eea
where we keep $\lambda$ as a book-keeping parameter. 
Ref.~\cite{Beneke:2002ni} gives implicit all-order in $\lambda$ relations 
between the $\xi$, $ A_c$ and $\hat\xi$, $\hat A_c$ fields in terms of a
soft Wilson line from $x_-^\mu$ to $x^\mu$ and a collinear Wilson line. 
Expanding these relations we obtain 
\bea\label{eq:FG1}
  F^{(1)}[\xi, A_c]&=&  ig_s x_\perp A_s(x_-)\xi\,,\nn\\
  G^{(1)\mu}[\xi, A_c] &=& [ig_s x_\perp A_s(x_-), A_c^\mu] + \frac12 n_+^\mu \left( W_c x_{\perp}^\nu n_{-\rho}  F_s^{\nu\rho}  W_c^\dag - x_{\perp}^\nu n_{-\rho}  F_s^{\nu\rho} \right)\,.
\eea
The field redefinition for $n_+ A_c$ is fixed only up to residual gauge 
transformation and the above relations hold only for a specific, 
non-covariant gauge choice. In practice, we want to use a covariant 
gauge fixing in the original fields to ensure that the SCET LSZ 
$Z$-factor of the collinear fields as well as the $\overline{\rm MS}$ 
field 
renormalization factor does not receive any $\mathcal{O}(\lambda)$ 
contribution at one-loop,\footnote{Since purely collinear interactions
are $\mathcal{O}(\lambda^0)$, power-suppressed contributions to the two-point function of collinear
fields can only arise via soft loops. In section 3.2.1 of \cite{Beneke:2018rbh} it was shown that, at one-loop
order, such contributions vanish, even in presence of an off-shell regulator.} such that the LSZ $Z$-factor remains the same with the 
above field redefinition. 
This choice allows to directly compare the on-shell Green functions computed with 
the original fields to those computed with the redefined fields.\footnote{On-shell 
S-matrix elements do not depend on specific 
fields used to construct the theory, however the on-shell amputated Green 
functions may be different for different sets of fields, if the field 
redefinition affects LSZ $Z$-factors.}
At leading power, this corresponds to the gauge fixing adopted in 
\cite{Beneke:2018rbh}. However, after expressing the gauge-fixing Lagrangian 
in terms of the new fields, new $\mathcal{O}(\lambda)$ collinear-soft gluon 
interactions appear. These do not contribute to diagrams with a single 
insertion of the $\mathcal{O}(\lambda)$ contribution to the gauge-fixing Lagrangian with
one or several collinear external particles, as will be considered below. 
For the detailed discussion of gauge invariance and the construction 
of the field redefinition we therefore refer to \cite{inprep}. 

We expand the full collinear Lagrangian in $\lambda$ in terms of the 
original fields, and alternatively in terms of the redefined fields,
\be
\hat {\cal L}[\hat \xi,\hat A_c] = \sum_n \lambda^n \hat {\cal L}^{(n)}[\hat \xi,\hat A_c],
\qquad
{\cal L}[\xi, A_c]  = \sum_n \lambda^n 
 {\cal L}^{(n)}[ \xi, A_c] \,.  
\ee
The terms $ {\cal L}^{(n)}$ are manifestly gauge invariant order by order in $\lambda$ \cite{Beneke:2002ni}.
Even though the individual terms differ, the sum over all terms satisfies
${\cal L}[\xi,A_c] = \hat{\cal L}[\hat\xi,\hat A_c]$.

For computing the off-shell regulated S-matrix elements, we need to
consider Green functions of the \emph{original} fields. 
Taking the $\lambda$-derivative of the appropriate generating functional  
with respect to the book-keeping parameter, and then formally
setting $\lambda=0$, generates the $\mathcal{O}(\lambda)$  time-ordered 
product insertion.  Besides, we take functional derivatives with respect to 
the appropriate sources to generate the insertion of the current ${\cal J}$ 
and the fields in the Green function of interest. Expressing the generating 
functional in terms of either the original or the redefined fields 
gives\footnote{For SCET current operators composed of gauge-invariant collinear building blocks evaluated
at $x=0$, the field redefinition becomes trivial. Therefore, there is no need to discriminate between ${\cal J}(0)$ and 
$\hat{\cal J}(0)$.} 
\bea\label{eq:WardLambda}
 &\langle 0|T\{ {\cal J}(0)\,\left(i\int_x{\hat {\cal L}}^{(1)}\right)\,\hat \xi(p_1)\cdots \hat \xi(p_n)\}|0\rangle 
 = \langle 0|T\{ {\cal J}(0)\,\left(i\int_x {\cal L}^{(1)}\right)\, \xi(p_1)\cdots \xi(p_n)\}|0\rangle &\nn\\
& + \sum_{r=1}^n\langle 0|T\{ {\cal J}(0) \xi(p_1)\cdots \xi(p_r)\cdots \xi(p_n)\}|0\rangle\Big|_{\xi(p_r)\to F^{(1)}(p_r)} \,.&
\eea
This relation expresses the content of the Ward identity for 
eom operators. It can be trivially extended to Green 
functions with gluon fields, conjugated collinear quark fields as well 
as additional fields in other collinear directions.

The left-hand side corresponds to the Green function that must be used to 
extract the anomalous dimension. Instead, it is more efficient to use 
the sum of terms on the right. The first term after the equality sign 
corresponds to diagrams with a single insertion of the manifestly 
gauge-invariant Lagrangian $ {\cal L}^{(1)}$, which vanishes  \cite{Beneke:2018rbh}.
The second line corresponds to replacing one of the fields that generate 
the external lines by the composite operator $F^{(1)}$. These terms yield 
the mixing of the time-ordered product into the $J^{A1}$ current. For Green 
functions with collinear gluons, one has to replace $F^{(1)}$ by 
$G^{(1)}$ for every gluon field appearing.


We now return to the explicit computation of mixing of the time-ordered 
products at $\mathcal{O}(\lambda)$ into the $J^{A1}$ currents, and generalize 
the previous treatment to the non-abelian case making use of the 
Ward identity \eqref{eq:WardLambda}. The Feynman rules for the 
contribution from the insertion of $F^{(1)}$ or $G^{(1)}$ for an 
external outgoing collinear antiquark or gluon, respectively,
to the off-shell regulated S-matrix element are given by 
\be
\begin{array}{lll}
\hspace*{-6.6cm}
  \raisebox{-13mm}{\includegraphics[width=0.2\textwidth]{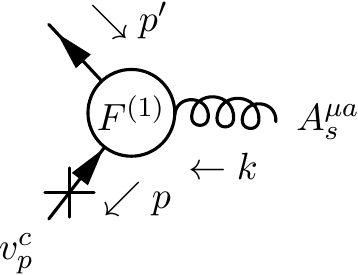}} &\hspace*{5mm}&
   ig_s t^a X_\perp^\mu \frac{p^2}{n_+p}\frac{\slashed{n}_+}{2}v_p^c\,,
\end{array}
\label{eq:Fvertex}
\ee

\be
\begin{array}{lll}
  \raisebox{-13mm}{\includegraphics[width=0.2\textwidth]{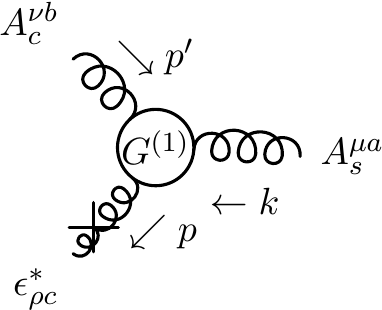}} &\hspace*{5mm}&
   g_s f^{abc} p^2 \left( X_\perp^\mu g^{\nu\rho} - \frac12 \frac{n_+^\nu n_+^\rho}{n_+p} X_\perp^\lambda n_-^\sigma (k_\lambda \delta^\mu_\sigma - k_\sigma \delta^\mu_\lambda)\right)\epsilon_{\rho c}^*\,.
\end{array}
\label{eq:Gvertex}
\ee
Here the cross denotes the external line, that is multiplied with the 
inverse free propagator and the external spinor or polarization vector, 
respectively, as appropriate for the off-shell regulated S-matrix element.
The symbol  $X_\perp^\mu$ denotes the momentum-derivative on the 
momentum-conservation delta-function, and is defined as in 
(A.12) of \cite{Beneke:2018rbh}.

Both vertices are proportional to $p^2$, i.e. they vanish in the 
on-shell limit unless this factor is cancelled by a factor $1/p^2$ coming 
from the diagram. This occurs for example in one particle reducible (1PR) diagrams with a loop in 
the external line, for which $F^{(1)}$ or $G^{(1)}$ are part of the loop. 
These are precisely the type of contributions that are cancelled by the
corresponding LSZ $Z$-factor related to the ``composite'' operator 
used to generate the external states when computing on-shell matrix 
elements. In the present case, these contributions vanish because when 
closing the soft gluon loop on the same collinear direction, the 
leading-power soft interaction yields a factor of $n_-^\mu$, which vanishes 
when connecting the soft gluon to $F^{(1)}$ and $G^{(1)}$.
This is in line with the observation that the $\mathcal{O}(\lambda)$ 
corrections to the LSZ Z-factor vanish in dimensional regularization and 
ensures that \emph{on-shell} Green functions of the original or 
redefined fields are identical in covariant gauges.  

Another possibility is to connect the soft gluon to another collinear 
direction. In this case, the momentum derivative contained in $X_\perp$ 
produces a soft loop which is non-vanishing, and contains a propagator 
with two powers as in (\ref{eq:Amp_SCET}), that can
lead to the factor $1/p^2$, which cancels the $p^2$ from the vertex. 
This yields the contributions that we obtained 
before from the explicit insertion of the eom Lagrangian.


\begin{figure}
\begin{center}
\includegraphics[width=0.5\textwidth]{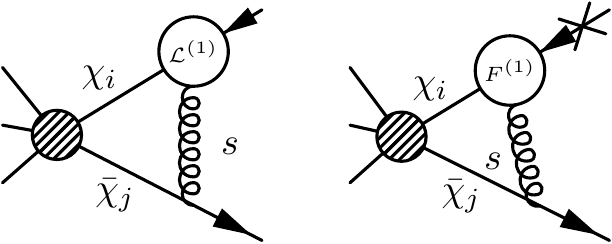}
\end{center}
\caption{\label{fig:soft_qq_F1}
Diagrams that describe the mixing $\bar\chi_j J_{\chi_i}^{T1}\to \bar\chi_j[i\partial_{\perp i}^\mu\chi_i]$ according to the Ward identity \eqref{eq:WardLambda}.}
\end{figure}

The relevant diagrams for the mixing 
$\bar\chi_j J_{\chi_i}^{T1}\to \bar\chi_j[i\partial_{\perp i}^\mu\chi_i]$ 
are shown in Figure~\ref{fig:soft_qq_F1}. The left diagram corresponds to 
the first line on right-hand side of \eqref{eq:WardLambda} and it 
vanishes \cite{Beneke:2018rbh}. The right diagram  contains the insertion of $F^{(1)}$, and yields
the divergence that can be absorbed by a counterterm proportional to the 
$J^{A1}$ operator. The corresponding contribution to the soft anomalous dimension 
equals \eqref{eq:NLPanomalousdim} obtained earlier.  This confirms that 
the method developed in this section is equivalent to the explicit use of 
the Lagrangian with original fields, that is, it checks the Ward identity 
(\ref{eq:WardLambda}), and shows that the previous  result 
\eqref{eq:NLPanomalousdim} is also valid for the non-abelian case.
Similarly, for the collinear gluon building block in the $i$ direction and 
a collinear quark in the other direction (that is, $j$ or $k$, whichever 
is different from $i$), which requires using $G^{(1)}$, we find 
\be\label{eq:gchiA}
 \gamma^{kj}_{P_iQ_i} = g_{\perp i}^{\rho\sigma}\,
\frac{\alpha_s}{\pi}\, \T_k\cdot \T_j\,
\frac{n_{j-}^\mu\delta_{k i}}{(n_{k-}n_{j-})n_{i+}p_i} + 
(k\leftrightarrow j)   \,,
\ee
with $P_i=J^{A0}_{1}\ldots J^{T1}_{{\cal A}_{\perp i}^\rho}\ldots J^{A0}_{N}$ 
and $Q_i=J^{A0}_{1} \ldots J^{A1}_{[i\partial_{\perp i}^\mu 
{\cal A}_{\perp i}^\sigma]}\ldots J^{A0}_{N}$. In \eqref{eq:gchiA}, we 
display explicitly the open Lorentz indices $\rho,\sigma,\mu$ 
of the gluon building blocks in the $i$ direction
for clarity, but leave implicit all open Lorentz or Dirac indices in the 
other collinear directions. Altogether, $\gamma^{kj}_{P_iQ_i}$ is diagonal 
in all these indices (including the $i$, $k$ and $j$ direction),
and \eqref{eq:gchiA} agrees with \eqref{eq:NLPanomalousdim} when leaving 
also the indices of the building block for the $i$ direction implicit, 
that is, when suppressing the factor $g_{\perp i}^{\rho\sigma}$.
Using this compact notation, we find that the anomalous dimension for the case of a gluon building block
in both the $k$ and the $j$ directions has also a form identical to \eqref{eq:NLPanomalousdim}.

\section{Collinear emission, mixing into B1 
currents and divergent convolutions}
\label{sec:collemission}

To check the consistency of the above results we consider the on-shell 
amplitude with an additional transverse $i$-collinear gluon, 
$\langle \bar q(p_1) q(q) g(p_2)| J(0) | 0\rangle$, in QCD and its 
representation in SCET. Here $J(0)=\bar{\psi}\,\Gamma\psi$ denotes 
a QCD quark-antiquark current, and $p_1, p_2$ are the $i$-collinear 
momenta of the anti-quark and gluon. 
To simplify the notation, we restrict ourselves in this 
section to two collinear directions, $i$ and $j$, and assume that the 
power-suppression of the operator arises in the $i$-direction. Then,  
at $\mathcal{O}(\lambda)$, the collinear fields for the $j$-direction 
always consist of the 
single $\bar\chi_j$. We therefore simply write $J^{B1\,\mu}$ for 
$\bar\chi_j\Gamma J^{B1\,\mu}_i = \bar\chi_j\Gamma \mathcal{A}_{\perp i}^\mu 
\chi_i$. Similarly, $J^{A1\,\mu}$, $J^{T1}$ etc. are understood to 
contain the factor $\bar\chi_j\Gamma$.

We first restrict ourselves to the 
abelian terms and focus on divergent terms of the form 
$1/p^2\times 1/\epsilon$, where $p^2=(p_1+p_2)^2$ is the virtuality 
of the collinear quark-gluon pair. We then find, as was the case for 
the amplitude without extra emission, that the 
sub-leading power eom Lagrangian  $\Delta{\cal L}^{(1)}_{\xi_i}$ 
(\ref{eq:delL1}) vanishes when inserted into the {\em on-shell} amplitude, 
since the soft loop is scaleless. However, the counterterm  
$\delta Z_{J^{T1},\,J^{A1\,\mu}}$ (\ref{eq:ZT1A1}) from mixing of 
$\Delta{\cal L}^{(1)}_{\xi_i}$ into the $J^{A1}$ operator contributes 
a $1/p^2\times 1/\epsilon$ pole from the one-particle reducible 
tree diagram with counterterm insertion. We find that this contribution 
is precisely what is required to reproduce the divergence of the full 
QCD amplitude in the limit $p^2\to 0$. 

The SCET calculation of this amplitude at $\mathcal{O}(\lambda)$ includes 
a term of the form 
\begin{equation}
\int_0^1 dx\,C^{(A0,B1)}(x) \,\langle \bar q(q) q(p_1) g(p_2)|
[\bar\chi_j \mathcal{A}_{\perp i}\chi_i](x)|0\rangle,
\label{eq:B1convolution}
\end{equation}
representing the convolution of the matrix element of a B1-type operator 
in momentum space with its hard QCD-to-SCET matching coefficient. The variable $x$ is the 
fraction of total $i$-collinear momentum carried by the gluon field in 
the operator. We find that the convolution of the tree-level 
coefficient function with the one-loop matrix element contributes 
to the above mentioned $1/p^2\times 1/\epsilon$ pole from the part of the 
coefficient function $C^{(A0,B1)}(x)$ which is proportional to $1/x$. 
However, when the convolution is performed after the matrix element 
is expanded in $\epsilon$, as would be the standard procedure, the 
convolution integral contains an unregulated divergence as 
$x\to 0$ and is ill-defined. Instead, the convolution must be done
before the matrix element is expanded in $\epsilon$, in which 
case one finds agreement with the full QCD calculation, as stated above.

\begin{figure}
\begin{center}
\includegraphics[width=0.85\textwidth]{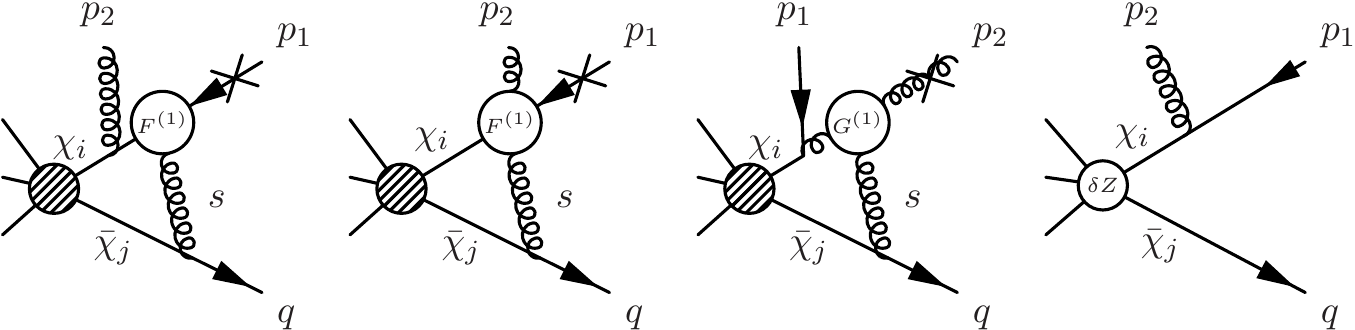}
\end{center}
\caption{\label{fig:soft_qqg_FG1}
Diagrams that describe a potential mixing 
$J^{T1}\to \bar\chi_j {\cal A}_{\perp i}^\mu\chi_i$. Momenta are outgoing. 
We omit diagrams with insertions of $ {\cal L}^{(1)}$,
because they vanish.}
\end{figure}

This observation has important implications for the soft mixing 
through the eom Lagrangian into B1 operators. To extract the anomalous 
dimension, we evaluate the matrix element of the time-ordered product operator 
in an external state with an additional, transverse $i$-collinear 
gluon as above, but now each leg carries a small off-shellness. 
We also return to the non-abelian theory. 
The relevant diagrams for soft mixing of 
$J^{T1}\to J^{B1\,\mu}=\bar\chi_j {\cal A}_{\perp i}^\mu\chi_i$ 
are shown in Figure~\ref{fig:soft_qqg_FG1}. Diagrams with insertions of 
${\cal L}^{(1)}$ are already omitted, because they 
vanish \cite{Beneke:2018rbh}. The 
insertions of the vertices $F^{(1)}$ and $G^{(1)}$ can appear only
at the external lines according to the Ward identity~\eqref{eq:WardLambda}. 
The second diagram in Figure~\ref{fig:soft_qqg_FG1} actually 
vanishes, because no collinear gauge field appears in $F^{(1)}$; see (\ref{eq:Fvertex}). 
The last diagram in Figure~\ref{fig:soft_qqg_FG1} is a 1PR tree-diagram with the 
insertion of the counterterm $\delta Z_{J^{T1},\,J^{A1\,\mu}}$ 
(\ref{eq:ZT1A1}) at the operator vertex. 

The soft loop diagrams are no longer scaleless in the presence of the 
off-shell regulator. The $1/\epsilon$ pole of the soft contribution is 
proportional to $1/p^2$  i.e. it  is \emph{non-local}. 
It cannot be absorbed into a counterterm constructed 
from the $\mathcal{O}(\lambda)$ SCET operators, which are local except 
for the direction of the light-cone of collinear fields. It is also not 
possible to interpret the divergence as mixing of non-local time-ordered 
products into themselves as they have non-vanishing tree-level matrix 
elements only for external states with soft particles. It follows that 
contrary to the mixing $J^{T1} \to J^{A1\,\mu} = 
\bar{\chi}_j i \partial_\perp^\mu \chi_i$ it is not possible to define 
consistently the corresponding mixing into the B1-operator
$\bar{\chi}_j \mathcal{A}_{\perp i}^\mu \chi_i$.
 
A related problem appears, however, for collinear loops. While the 
mixing of the B1 operator into itself is well-defined, and has been 
computed in \cite{Beneke:2017ztn} (see also \cite{Hill:2004if,Beneke:2005gs} for related computation in heavy-quark physics), some of the pole parts of the collinear 
contribution to the off-shell regulated matrix element are given by the 
convolution of the anomalous dimension with the tree-level 
matching coefficient similar to (\ref{eq:B1convolution}), which is 
again divergent. However, when the convolution is performed in $d$ 
dimensions before expanding the collinear loop integral in $\epsilon$, 
it generates a $1/\epsilon\times 1/p^2$ term, which exactly cancels 
the non-local divergence from the soft loops. Note that a similar 
cancellation of non-local $1/\epsilon\times \ln p^2$ poles between 
collinear and soft loops already appears in the computation of the 
one-loop cusp anomalous dimension (\ref{eq:LPanomalousdimexp}) at leading 
power. In that case, however, the cancellation appears within the  
anomalous dimension of the A0 operator alone, while at 
next-to-leading power the cancellation would have to happen between 
different entries of the anomalous dimension matrix, each of 
which is separately ill-defined.

To solve both problems, that is, to construct a well-defined anomalous 
dimension matrix including a consistent treatment of the divergent 
convolution, we modify the basis 
of $\mathcal{O}(\lambda)$ SCET objects. We first define the 
``singular'' B1 operator $J^{B1s\,\mu}$,
\bea\label{eq:B1s}
J^{B1s\,\mu}(t_i, t_j) &=& \bar\chi_j(t_jn_{j+}) \Gamma 
\left[\frac{1}{in_{i+}\partial}{\cal A}_{\perp i}^\mu(t_{i}n_{i+})\right]
\chi_i(t_{i}n_{i+})\,.
\eea
Both building blocks in the $i$-collinear direction are evaluated at 
the same position $t_i$. The inverse derivative, which operates only 
inside the square bracket, translates into a factor 
$1/x$, the momentum fraction carried by the gluon building block, 
such that the operator itself, and hence its anomalous dimension, 
absorbs the singular endpoint behaviour of the matching coefficient of 
the standard B1-type operator, when the momentum of the gluon becomes soft. 
We then trade the objects $J^{A1\,\mu}$, $J^{B1\,\mu}$ and $J^{T1}$ for 
\begin{equation}
\label{eq:Jcheck}
J^{A1\,\mu}, \quad J^{B1\,\mu}_{\rm reg}\,,\quad 
\check J \equiv J^{T1}+\frac{2n_{j-}^\mu}{n_{j-}n_{i-}}J^{B1s}_\mu, 
\end{equation}
where the ``regular'' B1 operator is defined by
analogy with the ``$+$ distribution'' as 
\be\label{eq:B1reg}
J_{\rm reg}^{B1\,\mu}(x) = 
\lim_{\eta\to 0^+}\left[\theta(x-\eta) J^{B1\,\mu}(x) - \eta \delta(x-\eta) 
\int_\eta^1 dz\frac{J^{B1\,\mu}(z)}{z} \right]\,.
\ee
The limit $\eta\to 0^+$ 
should be taken after the convolution integral with the coefficient function 
over $x$ is evaluated. 
The effect of the subtraction is to precisely remove the $1/x$ term 
from the coefficient function since 
this part is already included in the singular B1 operator. The 
operator $\check J$ is constructed in such a way that the non-local 
$1/\epsilon$ pole cancels in the sum of the time-ordered product 
and singular B1-operator term. The issue of a divergent convolution 
does not arise here since the singular B1 operator has the $i$-collinear 
fields at the same position. The originally divergent convolution 
is implicitly part of the operator, hence performed in $d$ dimensions, 
and the divergence subtracted by renormalization.
Several consistency requirements must be satisfied to make this 
construction valid. 

First, the two terms added together to $\check J$ 
must have the same divergence and the same hard matching coefficients 
to all orders in perturbation theory, so that $\check J$ renormalizes 
as a single object. Since the time-ordered product operator 
$J^{T1}$ inherits its evolution from the LP A0-operator, the
singular B1-operator must have the same anomalous dimension 
and coefficient function as $J^{A0}$. 
We have seen that this holds for the tree-level 
matching coefficients by construction and we checked at the one-loop 
level that the anomalous dimension for $\check J\to \check J$ mixing 
indeed agrees with the one of the leading-power current $J^{A0}$. 
In fact, the combination $\check J$ of time-ordered product 
and singular B1 operator is reparameterization invariant and, hence, 
should stay intact to all orders. 

Second, the anomalous dimension for the mixing of 
$J^{B1s} \to J^{B1}_{\rm reg}(y)$ must be finite as $y\to 0$. We 
find, for the operator with contracted colour indices 
($C_F=4/3, C_A=3$), 
\be
\gamma_{J^{B1s\,\mu},J^{B1\, \nu}_{\rm reg\, \alpha\beta}(y)} =  
\frac{\alpha_s}{4\pi}\frac{1}{n_{i+} (p_1+p_2)}\left(\Gamma\left[ 
(C_F-C_A/2)\,\gamma_{\perp i}^\mu\gamma_{\perp i}^\nu\,\frac{\ln(\bar y)}{y} 
+C_F\gamma_{\perp i}^\nu\gamma_{\perp i}^\mu\right]\right)_{\alpha\beta} \,,
\ee
which is indeed finite as $y\to 0$, and thus the mixing terms in the ADM 
for the new building block $\check J$ are 
\bea
  \gamma_{\check J,J^{A1\,\nu}} &=& \gamma_{J^{T1},J^{A1\, \nu}}\,,\nn\\
  \gamma_{\check J,J_{\rm reg \,\alpha\beta}^{B1\,\nu}(y)} &=& \frac{2n_{j-}^\mu}{n_{j-}n_{i-}}\gamma_{J^{B1s\, \mu},J_{\rm reg \, \alpha\beta}^{B1\,\nu}(y)}\, ,
\eea
where we used that $J^{B1s}$ does not mix into $J^{A1}$ and that $J^{T1}$ does not mix into $J^{B1}_{\rm reg}$.
The operator $J_{\rm reg}^{B1}(x)$ is 
renormalized by the same kernel as the unregularized 
$J^{B1}(x)$ operator~\cite{Beneke:2017ztn,Beneke:2018rbh}. This completes the 
mixing of $\mathcal{O}(\lambda)$ quark-antiquark (+ gluon) currents 
including the soft mixing from eom operators and the renormalization 
of divergent convolutions.

Third, and finally, the coefficient function of the unregularized B1 
operator must not be more singular than $1/x$ as $x\to 0$, and in particular, 
must not contain $1/x\times \ln^n x$ at higher orders, since this would 
invalidate the subtraction (\ref{eq:B1reg}) in the definition of the 
regular operator. That is, to all orders in $\alpha_s$, we must have 
$C^{B1}(x) = c(\alpha_s)/x + \mbox{less singular terms}$. 

The appearance of a divergent convolution with the hard 
coefficient is a manifestation 
of the breakdown of naive soft-collinear factorization in SCET. 
Indeed, the $1/x$ behaviour of the tree-level B1-operator 
matching coefficient appears when the gluon in the operator, 
assumed to be collinear, becomes soft. In full QCD, the $1/x$ 
behaviour originates from 
the coupling of the $i$-collinear gluon to {\em another} collinear 
direction. This explains why the cancellation of the non-local pole 
occurs between the singular part of the B1-type operator and the time-ordered 
product with soft gluon exchange between two collinear directions, 
and why only the sum $\check J$ is well-defined from the 
renormalization perspective. Divergent convolution integrals 
are familiar in SCET, but usually appear in the breakdown of 
factorization between modes of the same virtuality but different 
light-cone momentum or rapidity \cite{Beneke:2003pa}, resulting in a 
divergence that cannot be regulated dimensionally. 
Here, however, we encounter divergent convolutions with hard matching 
coefficients, related to the breakdown of factorization between 
regions of different virtuality, collinear and soft, 
which {\em are} regulated dimensionally. A similar problem has 
appeared first to our knowledge in the computation of 
electromagnetic corrections to $B_s\to \mu^+\mu^-$ 
\cite{Beneke:2017vpq}, 
where the computation was performed with an additional analytic 
regulator, since modes with different and equal virtuality 
are required, as well as in \cite{Alte:2018nbn} in a context 
with collinear and soft modes with different virtuality, which resembles more 
closely the situation discussed here. However, in both cases 
the endpoint divergence arose in a diagram with soft fermion 
exchange, whereas here it appears in a more standard context of 
soft-gluon exchange. We suspect that such divergences are generic 
for SCET beyond LP, at least as soon as the collinear directions 
are no longer back-to-back, which is always the case when more than two jets 
are involved.  It remains an open problem whether divergent convolutions 
in factorization formulas between functions of different virtuality 
can always be renormalized by the introducing singular and regular 
multi-jet operators as in the simplest case discussed above.

\section{Summary}

We investigated the mixing of time-ordered products of SCET $N$-jet 
operators with power-suppressed terms in the Lagrangian into 
$N$-jet operators and we found that the KSZ theorem is violated. That is, 
soft-collinear interaction Lagrangians that are equivalent by the 
equation of motion yield different anomalous dimensions for the 
$N$-jet operators. The violation is caused by the peculiar structure of 
interactions in SCET, involving the multipole expansion 
of soft fields that generates $x$-dependent terms in the Lagrangian, 
resulting in momentum derivatives, and by the fact that the 
anomalous dimension must capture a double $1/\epsilon^2$ pole per 
loop, which is related to the cusp anomalous dimension. The eom 
terms do not affect SCET on-shell matrix elements 
once the relevant counter\-terms have been obtained. 
However, the 
mixing of eom operators into physical operators implies that there 
is a preferred field representation for the anomalous dimension 
calculation, which is the one that reproduces correctly the IR 
singularities of QCD.

Notwithstanding this surprising result, SCET is a sensible EFT, 
since the eom terms and preferred set of fields are easily 
identified. Unlike other EFTs, the SCET Lagrangian is fixed completely 
by tree-level matching and no new operators are induced to any order 
in $\alpha_s$. Thus, even though eom operators mix into physical operators, 
only a finite number of such counterterms appears at a given order in 
the $\lambda$ expansion. The Lagrangian that produces the correct anomalous 
dimension is the one that reproduces the off-shell tree-level amplitudes 
in QCD. The additional soft mixing of eom operators must be added to 
the one-loop anomalous dimension computed 
in~\cite{Beneke:2017ztn,Beneke:2018rbh}, as illustrated here for the 
case of $\mathcal{O}(\lambda)$ power-suppressed quark-antiquark 
operators.

We confirmed that the additional soft mixing from eom Lagrangians into 
A1-type operators is actually required to reproduce IR divergences of 
QCD in SCET in more general kinematic situations. We checked this by 
performing the on-shell matching for A1 currents in the non-back-to-back 
kinematic configuration with non-vanishing transverse momenta with 
respect to the collinear direction axis. In addition, we checked that 
at one loop, the soft region of QCD with off-shell regulator is exactly 
reproduced at the integrand level by the sum of SCET physical and eom terms. 
We also find that the soft mixing is consistent with  
reparameterization invariance of SCET and is in fact required by 
Lorentz invariance of the cusp anomalous dimension. 

Extending the investigation to the mixing into B-type currents revealed 
divergent convolution integrals of the hard matching coefficient with 
the SCET matrix element for operators with more than one collinear 
field in a given direction. Such divergences have been observed before 
in a few cases involving soft fermion exchange 
\cite{Beneke:2017vpq,Alte:2018nbn}, but here they appear in a 
comparatively standard situation with soft gluons. They therefore 
seem to be generic and impede a consistent definition of the 
anomalous dimension matrix without further considerations. We proposed 
a split of the B-type operator into a singular and a regular part, 
such that the renormalization of the singular operator includes 
the consistent renormalization of the endpoint divergence of the 
convolution integral. The construction is supported by reparameterization invariance 
and passes the available consistency checks. The generality of the 
proposed solution should be further studied. The issue is of paramount 
importance for NLP resummation beyond the leading logarithms.  

Our analysis was formulated in the position-space 
formulation of SCET \cite{Beneke:2002ph,Beneke:2002ni}, but we expect 
the same results in the label formulation \cite{Bauer:2000yr,Bauer:2001yt}. 
Although there are no $x_\perp^\mu$ factors (hence momentum derivatives) 
in the label SCET Lagrangian, the propagator denominators with 
higher powers also appear in this formalism, as well as, evidently, 
the soft-collinear double $1/\epsilon^2$ poles. Momentum derivatives 
and $x$-factors also appear in the Lagrangian of potential 
non-relativistic field theory \cite{Pineda:1997bj,Beneke:1999zr}. 
Nevertheless, we do not expect a violation of the KSZ theorem 
in this case, despite the structural similarity of the 
$\mathcal{O}(\lambda)$ dipole interaction of the soft gauge 
field. In non-relativistic systems, the multipole expansion is 
performed in the spatial dimensions, rather than the transverse directions. 
After performing the integration over the time component of loop 
momentum, one obtains ordinary Feynman integrals in three dimensions, 
which can be regulated dimensionally. As a result one finds 
single poles in $\epsilon$ only and no convolution integrals 
that could be divergent.

\subsubsection*{Acknowledgements} We thank A.~Manohar for valuable 
comments. This work was supported in part 
by the BMBF grant  No.~05H18WOCA1.

\bibliography{paper}

\providecommand{\href}[2]{#2}\begingroup\raggedright\begin{thebibliography}{10}

\bibitem{Deans:1978wn}
W.~S. Deans and J.~A. Dixon, \emph{{Theory of Gauge Invariant Operators: Their
  Renormalization and S Matrix Elements}},
  \href{https://doi.org/10.1103/PhysRevD.18.1113}{\emph{Phys. Rev.} {\bfseries
  D18} (1978) 1113--1126}.

\bibitem{Politzer:1980me}
H.~D. Politzer, \emph{{Power Corrections at Short Distances}},
  \href{https://doi.org/10.1016/0550-3213(80)90172-8}{\emph{Nucl. Phys.}
  {\bfseries B172} (1980) 349--382}.

\bibitem{KlubergStern:1975hc}
H.~Kluberg-Stern and J.~B. Zuber, \emph{{Renormalization of Nonabelian Gauge
  Theories in a Background Field Gauge. 2. Gauge Invariant Operators}},
  \href{https://doi.org/10.1103/PhysRevD.12.3159}{\emph{Phys. Rev.} {\bfseries
  D12} (1975) 3159--3180}.

\bibitem{Joglekar:1975nu}
S.~D. Joglekar and B.~W. Lee, \emph{{General Theory of Renormalization of Gauge
  Invariant Operators}},
  \href{https://doi.org/10.1016/0003-4916(76)90225-6}{\emph{Annals Phys.}
  {\bfseries 97} (1976) 160}.

\bibitem{Espriu:1983zz}
D.~Espriu, \emph{{Renormalization of Gauge Invariant Operators and the Axial
  Anomaly}}, \href{https://doi.org/10.1103/PhysRevD.28.349}{\emph{Phys. Rev.}
  {\bfseries D28} (1983) 349}.

\bibitem{Collins:1994ee}
J.~C. Collins and R.~J. Scalise, \emph{{The Renormalization of composite
  operators in Yang-Mills theories using general covariant gauge}},
  \href{https://doi.org/10.1103/PhysRevD.50.4117}{\emph{Phys. Rev.} {\bfseries
  D50} (1994) 4117--4136},
  [\href{https://arxiv.org/abs/hep-ph/9403231}{{\ttfamily hep-ph/9403231}}].

\bibitem{Manohar:2018aog}
A.~V. Manohar, \emph{{Introduction to Effective Field Theories}},  in
  \emph{{Les Houches summer school: EFT in Particle Physics and Cosmology Les
  Houches, Chamonix Valley, France, July 3-28, 2017}}, 2018,
  \href{https://arxiv.org/abs/1804.05863}{{\ttfamily 1804.05863}}.

\bibitem{Criado:2018sdb}
J.~C. Criado and M.~P{\'{e}}rez-Victoria, \emph{{Field redefinitions in
  effective theories at higher orders}},
  \href{https://doi.org/10.1007/JHEP03(2019)038}{\emph{JHEP} {\bfseries 03}
  (2019) 038}, [\href{https://arxiv.org/abs/1811.09413}{{\ttfamily
  1811.09413}}].

\bibitem{Gambino:2003zm}
P.~Gambino, M.~Gorbahn and U.~Haisch, \emph{{Anomalous dimension matrix for
  radiative and rare semileptonic B decays up to three loops}},
  \href{https://doi.org/10.1016/j.nuclphysb.2003.09.024}{\emph{Nucl. Phys.}
  {\bfseries B673} (2003) 238--262},
  [\href{https://arxiv.org/abs/hep-ph/0306079}{{\ttfamily hep-ph/0306079}}].

\bibitem{Bauer:2000yr}
C.~W. Bauer, S.~Fleming, D.~Pirjol and I.~W. Stewart, \emph{{An Effective field
  theory for collinear and soft gluons: Heavy to light decays}},
  \href{https://doi.org/10.1103/PhysRevD.63.114020}{\emph{Phys. Rev.}
  {\bfseries D63} (2001) 114020},
  [\href{https://arxiv.org/abs/hep-ph/0011336}{{\ttfamily hep-ph/0011336}}].

\bibitem{Bauer:2001yt}
C.~W. Bauer, D.~Pirjol and I.~W. Stewart, \emph{{Soft collinear factorization
  in effective field theory}},
  \href{https://doi.org/10.1103/PhysRevD.65.054022}{\emph{Phys. Rev.}
  {\bfseries D65} (2002) 054022},
  [\href{https://arxiv.org/abs/hep-ph/0109045}{{\ttfamily hep-ph/0109045}}].

\bibitem{Beneke:2002ph}
M.~Beneke, A.~P. Chapovsky, M.~Diehl and T.~Feldmann, \emph{{Soft collinear
  effective theory and heavy to light currents beyond leading power}},
  \href{https://doi.org/10.1016/S0550-3213(02)00687-9}{\emph{Nucl. Phys.}
  {\bfseries B643} (2002) 431--476},
  [\href{https://arxiv.org/abs/hep-ph/0206152}{{\ttfamily hep-ph/0206152}}].

\bibitem{Beneke:2002ni}
M.~Beneke and T.~Feldmann, \emph{{Multipole expanded soft collinear effective
  theory with non-abelian gauge symmetry}},
  \href{https://doi.org/10.1016/S0370-2693(02)03204-5}{\emph{Phys. Lett.}
  {\bfseries B553} (2003) 267--276},
  [\href{https://arxiv.org/abs/hep-ph/0211358}{{\ttfamily hep-ph/0211358}}].

\bibitem{Becher:2009cu}
T.~Becher and M.~Neubert, \emph{{Infrared singularities of scattering
  amplitudes in perturbative QCD}},
  \href{https://doi.org/10.1103/PhysRevLett.102.162001,
  10.1103/PhysRevLett.111.199905}{\emph{Phys. Rev. Lett.} {\bfseries 102}
  (2009) 162001}, [\href{https://arxiv.org/abs/0901.0722}{{\ttfamily
  0901.0722}}].

\bibitem{Beneke:2004in}
M.~Beneke, F.~Campanario, T.~Mannel and B.~D. Pecjak, \emph{{Power corrections
  to $\bar{B} \to X_u \ell \bar{\nu} \,(X_s \gamma)$ decay spectra in the
  `shape-function' region}},
  \href{https://doi.org/10.1088/1126-6708/2005/06/071}{\emph{JHEP} {\bfseries
  06} (2005) 071}, [\href{https://arxiv.org/abs/hep-ph/0411395}{{\ttfamily
  hep-ph/0411395}}].

\bibitem{Larkoski:2014bxa}
A.~J. Larkoski, D.~Neill and I.~W. Stewart, \emph{{Soft Theorems from Effective
  Field Theory}}, \href{https://doi.org/10.1007/JHEP06(2015)077}{\emph{JHEP}
  {\bfseries 06} (2015) 077},
  [\href{https://arxiv.org/abs/1412.3108}{{\ttfamily 1412.3108}}].

\bibitem{Freedman:2014uta}
S.~M. Freedman and R.~Goerke, \emph{{Renormalization of Subleading Dijet
  Operators in Soft-Collinear Effective Theory}},
  \href{https://doi.org/10.1103/PhysRevD.90.114010}{\emph{Phys. Rev.}
  {\bfseries D90} (2014) 114010},
  [\href{https://arxiv.org/abs/1408.6240}{{\ttfamily 1408.6240}}].

\bibitem{Kolodrubetz:2016uim}
D.~W. Kolodrubetz, I.~Moult and I.~W. Stewart, \emph{{Building Blocks for
  Subleading Helicity Operators}},
  \href{https://doi.org/10.1007/JHEP05(2016)139}{\emph{JHEP} {\bfseries 05}
  (2016) 139}, [\href{https://arxiv.org/abs/1601.02607}{{\ttfamily
  1601.02607}}].

\bibitem{Feige:2017zci}
I.~Feige, D.~W. Kolodrubetz, I.~Moult and I.~W. Stewart, \emph{{A Complete
  Basis of Helicity Operators for Subleading Factorization}},
  \href{https://doi.org/10.1007/JHEP11(2017)142}{\emph{JHEP} {\bfseries 11}
  (2017) 142}, [\href{https://arxiv.org/abs/1703.03411}{{\ttfamily
  1703.03411}}].

\bibitem{Moult:2017rpl}
I.~Moult, I.~W. Stewart and G.~Vita, \emph{{A subleading operator basis and
  matching for gg $\to$ H}},
  \href{https://doi.org/10.1007/JHEP07(2017)067}{\emph{JHEP} {\bfseries 07}
  (2017) 067}, [\href{https://arxiv.org/abs/1703.03408}{{\ttfamily
  1703.03408}}].

\bibitem{Beneke:2017mmf}
M.~Beneke, M.~Garny, R.~Szafron and J.~Wang, \emph{{Subleading-power $N$-jet
  operators and the LBK amplitude in SCET}},  in \emph{{13th International
  Symposium on Radiative Corrections: Application of Quantum Field Theory to
  Phenomenology (RADCOR 2017) St. Gilgen, Austria, September 24-29, 2017}},
  2017, \href{https://arxiv.org/abs/1712.07462}{{\ttfamily 1712.07462}}.

\bibitem{Beneke:2017ztn}
M.~Beneke, M.~Garny, R.~Szafron and J.~Wang, \emph{{Anomalous dimension of
  subleading-power N-jet operators}},
  \href{https://doi.org/10.1007/JHEP03(2018)001}{\emph{JHEP} {\bfseries 03}
  (2018) 001}, [\href{https://arxiv.org/abs/1712.04416}{{\ttfamily
  1712.04416}}].

\bibitem{Beneke:2018rbh}
M.~Beneke, M.~Garny, R.~Szafron and J.~Wang, \emph{{Anomalous dimension of
  subleading-power $N$-jet operators. Part II}},
  \href{https://doi.org/10.1007/JHEP11(2018)112}{\emph{JHEP} {\bfseries 11}
  (2018) 112}, [\href{https://arxiv.org/abs/1808.04742}{{\ttfamily
  1808.04742}}].

\bibitem{Moult:2018jjd}
I.~Moult, I.~W. Stewart, G.~Vita and H.~X. Zhu, \emph{{First Subleading Power
  Resummation for Event Shapes}},
  \href{https://doi.org/10.1007/JHEP08(2018)013}{\emph{JHEP} {\bfseries 08}
  (2018) 013}, [\href{https://arxiv.org/abs/1804.04665}{{\ttfamily
  1804.04665}}].

\bibitem{Beneke:2018gvs}
M.~Beneke, A.~Broggio, M.~Garny, S.~Jaskiewicz, R.~Szafron, L.~Vernazza et~al.,
  \emph{{Leading-logarithmic threshold resummation of the Drell-Yan process at
  next-to-leading power}},
  \href{https://doi.org/10.1007/JHEP03(2019)043}{\emph{JHEP} {\bfseries 03}
  (2019) 043}, [\href{https://arxiv.org/abs/1809.10631}{{\ttfamily
  1809.10631}}].

\bibitem{Ebert:2018gsn}
M.~A. Ebert, I.~Moult, I.~W. Stewart, F.~J. Tackmann, G.~Vita and H.~X. Zhu,
  \emph{{Subleading power rapidity divergences and power corrections for
  q$_{T}$}}, \href{https://doi.org/10.1007/JHEP04(2019)123}{\emph{JHEP}
  {\bfseries 04} (2019) 123},
  [\href{https://arxiv.org/abs/1812.08189}{{\ttfamily 1812.08189}}].

\bibitem{Beneke:2004rc}
M.~Beneke, Y.~Kiyo and D.~s. Yang, \emph{{Loop corrections to subleading heavy
  quark currents in SCET}},
  \href{https://doi.org/10.1016/j.nuclphysb.2004.05.018}{\emph{Nucl. Phys.}
  {\bfseries B692} (2004) 232--248},
  [\href{https://arxiv.org/abs/hep-ph/0402241}{{\ttfamily hep-ph/0402241}}].

\bibitem{Hill:2004if}
R.~J. Hill, T.~Becher, S.~J. Lee and M.~Neubert, \emph{{Sudakov resummation for
  subleading SCET currents and heavy-to-light form-factors}},
  \href{https://doi.org/10.1088/1126-6708/2004/07/081}{\emph{JHEP} {\bfseries
  07} (2004) 081}, [\href{https://arxiv.org/abs/hep-ph/0404217}{{\ttfamily
  hep-ph/0404217}}].

\bibitem{Beneke:2005gs}
M.~Beneke and D.~Yang, \emph{{Heavy-to-light B meson form-factors at large
  recoil energy: Spectator-scattering corrections}},
  \href{https://doi.org/10.1016/j.nuclphysb.2005.11.027}{\emph{Nucl. Phys.}
  {\bfseries B736} (2006) 34--81},
  [\href{https://arxiv.org/abs/hep-ph/0508250}{{\ttfamily hep-ph/0508250}}].

\bibitem{Beneke:2017vpq}
M.~Beneke, C.~Bobeth and R.~Szafron, \emph{{Enhanced electromagnetic correction
  to the rare $B$-meson decay $B_{s,d} \to \mu^+ \mu^-$}},
  \href{https://doi.org/10.1103/PhysRevLett.120.011801}{\emph{Phys. Rev. Lett.}
  {\bfseries 120} (2018) 011801},
  [\href{https://arxiv.org/abs/1708.09152}{{\ttfamily 1708.09152}}].

\bibitem{Alte:2018nbn}
S.~Alte, M.~K{\"o}nig and M.~Neubert, \emph{{Effective Field Theory after a
  New-Physics Discovery}},
  \href{https://doi.org/10.1007/JHEP08(2018)095}{\emph{JHEP} {\bfseries 08}
  (2018) 095}, [\href{https://arxiv.org/abs/1806.01278}{{\ttfamily
  1806.01278}}].

\bibitem{inprep}
M.~Beneke, M.~Garny, R.~Szafron and J.~Wang, \emph{{in preparation}}.

\bibitem{Catani:1998bh}
S.~Catani, \emph{{The Singular behavior of QCD amplitudes at two loop order}},
  \href{https://doi.org/10.1016/S0370-2693(98)00332-3}{\emph{Phys. Lett.}
  {\bfseries B427} (1998) 161--171},
  [\href{https://arxiv.org/abs/hep-ph/9802439}{{\ttfamily hep-ph/9802439}}].

\bibitem{Manohar:2002fd}
A.~V. Manohar, T.~Mehen, D.~Pirjol and I.~W. Stewart, \emph{{Reparameterization
  invariance for collinear operators}},
  \href{https://doi.org/10.1016/S0370-2693(02)02029-4}{\emph{Phys. Lett.}
  {\bfseries B539} (2002) 59--66},
  [\href{https://arxiv.org/abs/hep-ph/0204229}{{\ttfamily hep-ph/0204229}}].

\bibitem{Beneke:1997zp}
M.~Beneke and V.~A. Smirnov, \emph{{Asymptotic expansion of Feynman integrals
  near threshold}},
  \href{https://doi.org/10.1016/S0550-3213(98)00138-2}{\emph{Nucl. Phys.}
  {\bfseries B522} (1998) 321--344},
  [\href{https://arxiv.org/abs/hep-ph/9711391}{{\ttfamily hep-ph/9711391}}].

\bibitem{Beneke:2003pa}
M.~Beneke and T.~Feldmann, \emph{{Factorization of heavy to light form-factors
  in soft collinear effective theory}},
  \href{https://doi.org/10.1016/j.nuclphysb.2004.02.033}{\emph{Nucl. Phys.}
  {\bfseries B685} (2004) 249--296},
  [\href{https://arxiv.org/abs/hep-ph/0311335}{{\ttfamily hep-ph/0311335}}].

\bibitem{Pineda:1997bj}
A.~Pineda and J.~Soto, \emph{Effective field theory for ultrasoft momenta in
  {NRQCD} and {NRQED}}, {\emph{Nucl. Phys. Proc. Suppl.} {\bfseries 64} (1998)
  428--432}, [\href{https://arxiv.org/abs/hep-ph/9707481}{{\ttfamily
  hep-ph/9707481}}].

\bibitem{Beneke:1999zr}
M.~Beneke, \emph{{Perturbative heavy quark - anti-quark systems}},  in
  \emph{{Proceedings of the 8th International Symposium on Heavy Flavor Physics
  (Heavy Flavors 8)}}, 1999,
  \href{https://arxiv.org/abs/hep-ph/9911490}{{\ttfamily hep-ph/9911490}}.

\end{thebibliography}\endgroup

\end{document}